# CyberEye: Obtaining Data from Virtual Desktop by Video

**Bin Wei**

**Individual Research Lab**

**Email:** Bin.will@outlook.com

**Code&Demo Video: https://github.com/bin-will/cybereye**

**Abstract：**

**VDI is no longer safe and reliable anymore. VDI(Virtual Desktop Infrastructure, also called Cloud Desktop) is being widely used as working interface to avoid data exfiltration. With VDI client, end users can access internal data without obtaining data acturally.**

**In this paper, we present a new approach named CyberEye, to extract data from VDI by video even data transmission has been forbidden. By encoding data file to video, playing it in VDI meanwhile recording it in host PC, we can get full information of the data with video format, then decode it to recover the original data file.**

**The proof-of-concept on Citrix Workspace and servral other remote virtual desktops has strongly been proved the availability and reliability of the CyberEye. We introduce the usage in operation model to show how it's been designed and implemented in technical work section. And also, we have opened the source code to researchers for reproducing the work.**

# 1.Introduction

For most of enterprises, network security is controlled by firewall, antivirus and vulnerability scanner, mainly focusing on viruses, Trojan, vulnerabilities and sensitive operations. Application systems control data access policy normally. SA(Security Administrator) will set policies by roles and priviledges for end users to access different data. But these policies can not limit the usage by end users since data has been downloaded. Once the authorized user downloaded the data into his PC, it becomes much more difficult to control the spread scope of the data, therefore, that makes high risk of data exfiltration.

VDI is considered as a standard solution to prevent data exfiltration, widely used in many enterprises. By installed VDI client in host PC, end users will be able to connect intranet remotely via internet, and access internal data compliantly. Figure 1 shows general Citrix VDI architecture.

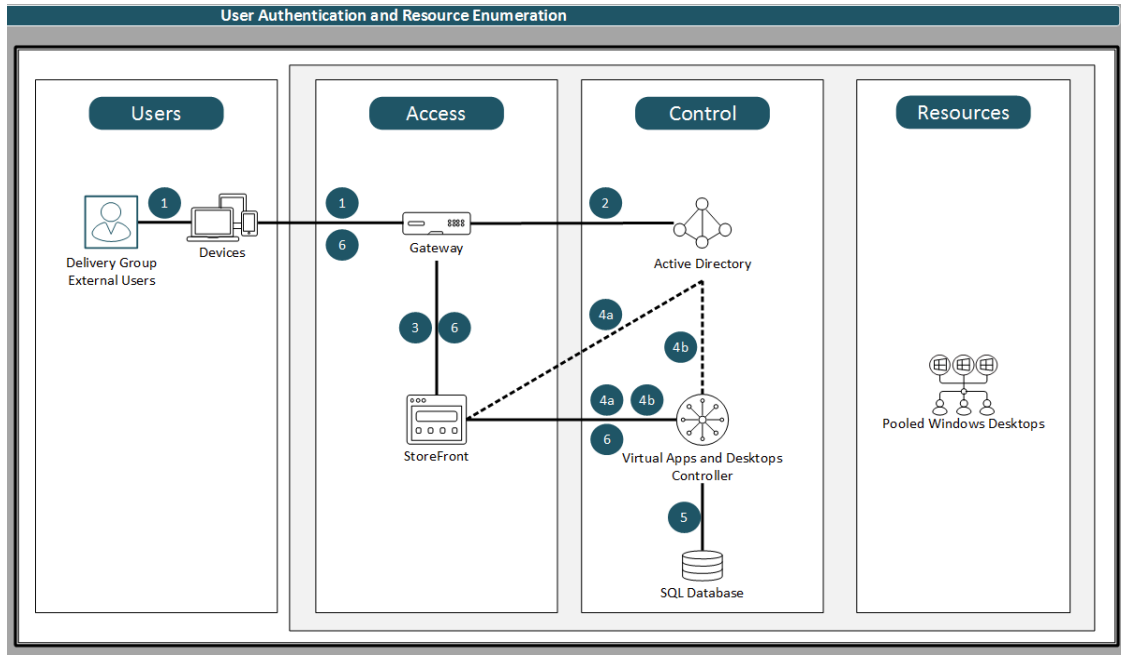

Fig. 1: Citrix VDI Architecture. 1). A user initiates a connection to the Gateway URL and provides logon credentials; 2). The credentials are validated against Active Directory. 3). Gateway forwards the user credentials to StoreFront. 4). StoreFront validates the user credentials by controller. 5). The Virtual Apps & Desktops delivery controller retrieves a list of available resources by querying the SQL Database. 6). The list of available resources is sent to StoreFront, which populates the user's Citrix Workspace app, Windows Start Menu or browser. Above graph and texts are from https://docs.citrix.com/en-us/tech-zone/learn/downloads/diagrams-posters_virtual-apps-and-desktops_poster.png.

As figure 2 shown, end users can use host PC with low security level to access internal data remotely through VDI, without obtaining the data actually. VDI will control data distribution policy, normally, permit restricted data input, like reading files from host PC hard drives, but deny data output like file writing, editing or transmiting to host PC drive, thus internal data will always be limited within VDI in data center, one way come in but no way go out.

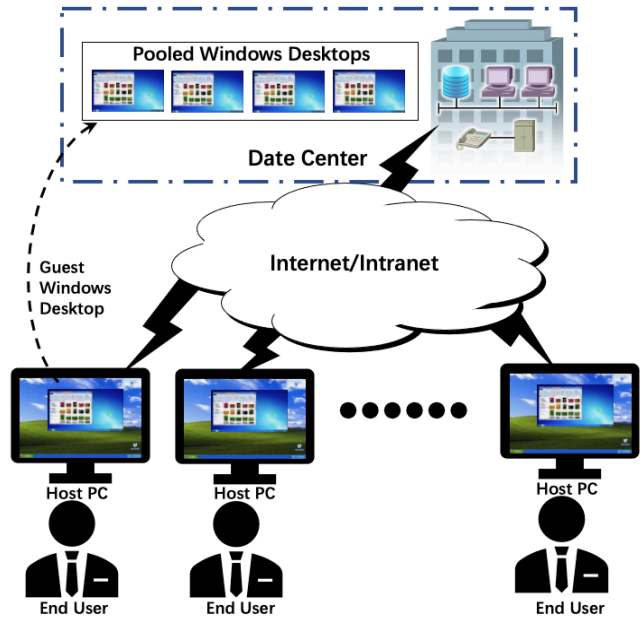

Fig. 2: End user can access internal data remotely via pooled desktops with his host PC.

Technically, data exfiltration still exits such as a quick peep or screen shooting, but has its strict limitations: : the data such as text must be visualized and recognized by human vision, or not required to be identical with the original file, such as image. Fortunately, most of data files are incompatible with these requirements, like binary files, zip files, etc., which will cause file damaged and invalidatation if only one bit got missed. Therefore, VDI is considered as a standard solution to prevent data exfiltration, widely used in many enterprises.

There still exist some innovative data exfiltration approaches from some researchers' pervious works, ultrasonic wave can be used to exfiltrate data between computers and smartphones [1][2], monitor's LED indicator can also leak data to smartphone's camara [3]. Inspired by their works, we've completed our research.

**Our Contribution:**

In this paper, we present a new data transmission approach and tool named CyberEye and open the source code, that can extract all types of data file precisely from VDI in spite of the limitations. This approach is tested on Citrix(https://www.citrix.com/) and servral other virtual desktop applications, but should work in any types of virtual desktop in theory. There is no additional special priviledges required, no sensitive behaviors conducted, no security alarms trggered.

The main steps: Upload CyberEye to virtual desktop, encode target file to video file, play it in virtual desktop meanwhile recording it in host PC, then decode recorded video to recover the original target file.

The rest of the paper is organized as follows：The operation model is discussed in Section 2. Technical work is presented in Section 3, including the design details about data encoding, transmission and decoding. Section 4 presents the evaluation and Section 5 is the conclusion.

# 2. Operation Model

Only four steps to extract target file out of virtual desktop in a simple and quiet way. As Figure 3, first, uploading CyberEye to VD(virtual desktop). Second, encoding target file to video file. Third, playing the video file within virtual desktop while recording screen on host PC. Forth, decoding the recorded video to original target file.

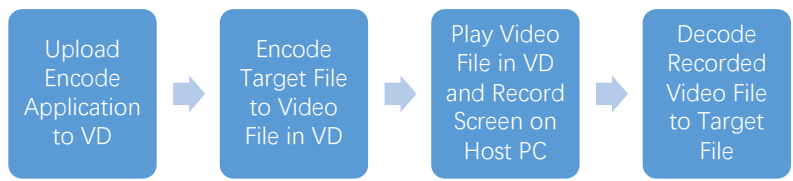

Fig. 3: Operation Model: 4 Steps to Get Data File

We will introduce the details of each operation step in the following contents to show how it works in real world.

## 2.1  Uploading Encode Tool to Virtual Desktop

We have opened the source code developed with Python3 and executable version supporting Win10/11/2008.

VDI supports disk mapping between host PC and VDI, like mapping drive D in Host PC to drive E in virtual desktop. User can access files in drive E to input files from host PC. Normally, SA(Security Admin) will set policy to permit READ ONLY but deny WRITE to host drive. Data will be one way trip from host PC to virtual desktop. Note that we focus on VDI in this paper and ignore other applications like email, website, im, etc., which should be controlled by other security system.

The weak host PC can upload CyberEye easily, then copy to virtual desktop through mapping drive. All of the operations are compliance, no installation, no virus, no scanning, no trojan, and no warning.

## 2.2　Encoding Target File to Video in Virtual Desktop

CyberEye is developed by Python3 with dependency OpenCV and Numpy, we have packaged executable Windows version by Pyinstaller. It can be executed without installation.

CyberEye runs on command line, just open a CLI terminal and assign target file, which can be any types of file. After seconds, it will be encoded to a MP4 video file which contains all information of the target file.

## 2.3　Playing Video in Virtual Desktop and Record Screen in Host PC

Since VDI blocks data outflow, we do not need to pull out the video file, in stead, we just play the video in virtual desktop with media player or any other players. Normally, media player is enough in all windows systems, no additional player required.

Any screen recorder application supporting region selection can be used in this step, we have tested apower(https://www.apowersoft.cn) in Windows and QuickTime in macOS, all work fine.

When starting play, lauch the recorder application quickly in host PC and select the player region to record, then wait until it finished, we will get the encoded video file containing all information from the target file. The video looks like many continuous QR codes.

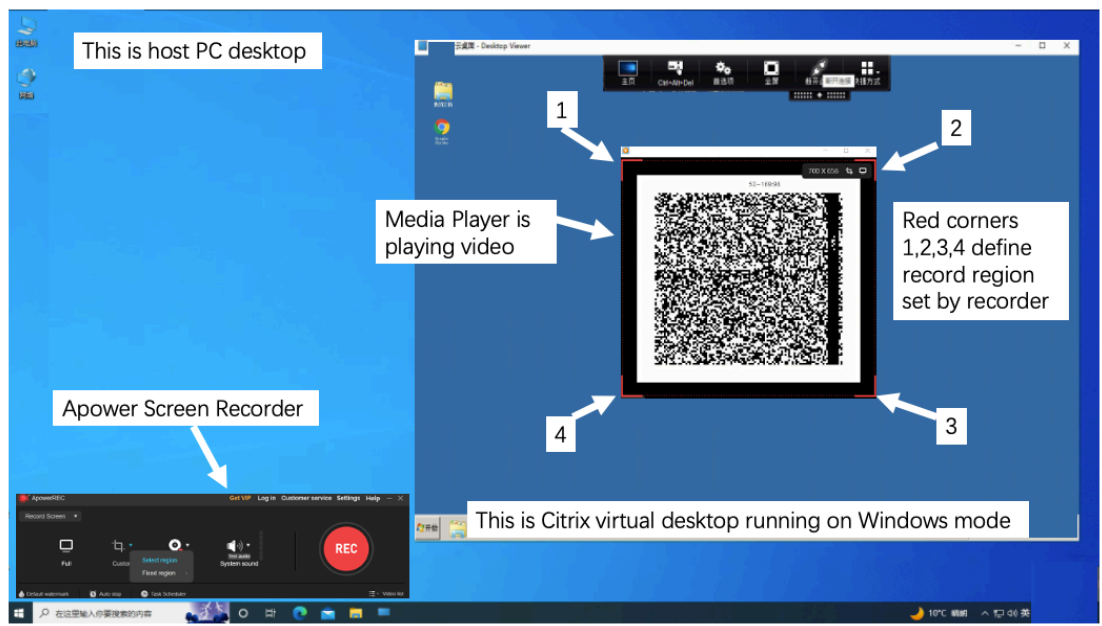

Fig. 4: Play video in virtual desktop and record in host PC desktop by Apower screen recorder. Note some icons were masked for privacy reason.

## 2.4 Decoding Recorded Video File to Target File

Once we get the video with full information, which is already out of the virtual desktop, the last step is to decode it with CyberEye. CyberEye is a full functional command line tool on encoding and decoding, just open a terminal and input the recorded video file, after a few seconds or minutes depending on the video size, we can recover the original data file.

# 3. Technical Works

In this section, we will introduce the design and implementation of CyberEye. We developed it with Python3, mainly depending on OpenCV and Numpy. OpenCV is a polular computer vision library, here works on reading and writing video file. Numpy is a powerful mathematics library, here works on operating data array.

The main motivation is ANY file in OS can be represented with 0-1 bits flow, as shown in figure 5. If we visualize 0 as a black square and 1 as a white square, one file can be encoded to a white-black square sequence which can be reshaped to frames, then can be combined to a video.

By reversing the process, we can decode the video and get original file. We will introduce each step below.

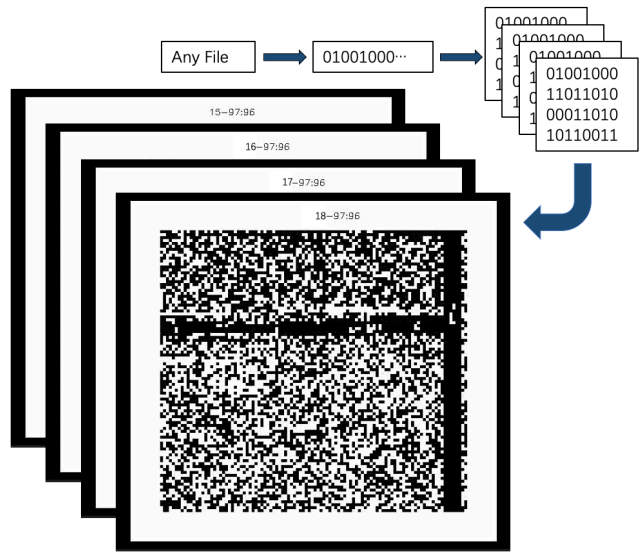

Fig. 5: Visualize a file to white-black square frames

## 3.1  Encoding Process

### a)  Reading File to Bits Sequence

Any files in OS are bits sequence represented with 0 and 1, any hex viewer like UltraEdit, HexDump can open it like figure 6.

```
00000480  00 01 04 01 03 02 04 02  05 07 06 08 05 03 0c 33  |...............3|
00000490  01 00 02 11 03 04 21 12  31 05 41 51 61 13 22 71  |......!.1.AQa."q|
000004a0  81 32 06 14 91 a1 b1 42  23 24 15 52 c1 62 33 34  |.2.....B#$.R.b34|
000004b0  72 82 d1 43 07 25 92 53  f0 e1 f1 63 73 35 16 a2  |r..C.%.S...cs5..|
000004c0  b2 83 26 44 93 54 64 45  c2 a3 74 36 17 d2 55 e2  |..&D.TdE..t6..U.|
000004d0  65 f2 b3 84 c3 d3 75 e3  f3 46 27 94 a4 85 b4 95  |e.....u..F'.....|
000004e0  c4 d4 e4 f4 a5 b5 c5 d5  e5 f5 56 66 76 86 96 a6  |..........Vfv...|
000004f0  b6 c6 d6 e6 f6 37 47 57  67 77 87 97 a7 b7 c7 d7  |.....7GWgw......|
00000500  e7 f7 11 00 02 02 01 02  04 04 03 04 05 06 07 07  |................|
00000510  06 05 35 01 00 02 11 03  21 31 12 04 41 51 61 71  |..5.....!1..AQaq|
00000520  22 13 05 32 81 91 14 a1  b1 42 23 c1 52 d1 f0 33  |"..2.....B#.R..3|
00000530  24 62 e1 72 82 92 43 53  15 63 73 34 f1 25 06 16  |$b.r..CS.cs4.%..|
00000540  a2 b2 83 07 26 35 c2 d2  44 93 54 a3 17 64 45 55  |....&5..D.T..dEU|
00000550  36 74 65 e2 f2 b3 84 c3  d3 75 e3 f3 46 94 a4 85  |6te......u..F...|
00000560  b4 95 c4 d4 e4 f4 a5 b5  c5 d5 e5 f5 56 66 76 86  |............Vfv.|
00000570  96 a6 b6 c6 d6 e6 f6 27  37 47 57 67 77 87 97 a7  |.......'7GWgw...|
00000580  b7 c7 ff da 00 0c 03 01  00 02 11 03 11 00 3f 00  |..............?.|
00000590  f5 54 92 49 25 29 24 92  49 4c 43 18 0c 86 80 7c  |.T.I%)$.ILC....||
000005a0  61 3a 74 92 52 d0 12 84  e9 24 a5 24 92 49 29 49  |a:t.R....$.$.I)I|
000005b0  24 92 4a 52 49 24 92 9f  ff d0 f5 54 92 49 25 29  |$.JRI$.....T.I%)|
```

Fig. 6: Part of sample image file in Hex Viewer

Focusing on main contents in middle of figure 6, 0~9 and a~f are hex representation of the file,

any file looks similar. As one Hex character can be transferred to eight binary bits, like hex e3 →
binary 1110 0011, we can read the target file and transfer to bits sequence easily by Python code.

## b) Constructing the Data Array

One video object is a frame sequence, one frame is a 2 Dimension (gray) or 3 Dimension (RGB) array with fixed width and height. In this paper, we choose 2D array to represent each frame simply, so the video object can be considered as 3D array since video object is frames sequence, but it is easy to update to RGB frame to contain more information in further works.

A bits sequence can be considered as 1D array, we can reshape it to 3D array easily by Numpy in servral lines code, and fill the blank of tail with value 128. Here, we define value 128 as END sigal of the real file data. In decoding process, we will mark the file end once we meet value 128.

Next step, update all value 1 in array to value 255, we can construct data array fill of 0, 255 and some 128 in the end. Here value 0 will be black pixel, 255 will be white pixel, 128 will be gray pixel in each frame of video. Here is a sample constructed video: https://github.com/bin-will/cybereye/blob/main/test.docx.mp4.

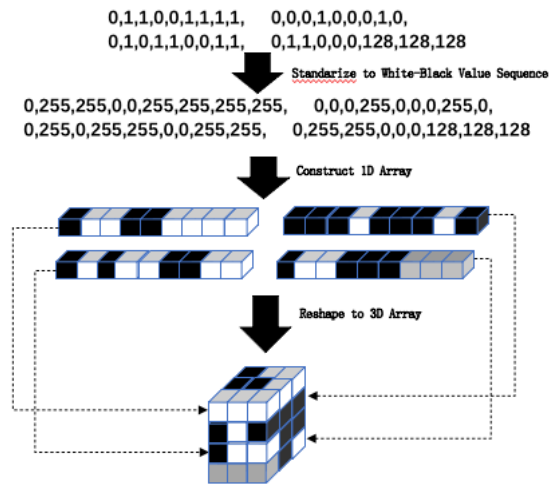

Fig. 7: Convert bits sequence to 3D array, each small cube presents one bit. Note the width and height of 3D array is 3x3 to simplify this figure, we choose 96x96 in real program.

## c) Constructing the Label Array

It is very important to add labels to each frame in video for the robust reasons:

i. Different video FPS between original video in virtual desktop and recorded video in host PC, orginal video is set to default 5 FPS, recorded video is set to default 25~30 FPS in Windows and 60 FPS in MacOS, some frames are duplicated.

ii. Network jam may cause frames lost or damaged, data maybe lost too.

Therefore, we must construct the label array corresponding with data array. We reserve 8 labels for each data frame in data array, all lable values will be transferred to value 0, 1 bits and then value 0, 255 also.

Table 1: Label defination in label array

| Label # | Defination | Comments |
| --- | --- | --- |
| 0 | None | Reserve |
| 1 | None | Reserve |
| 2 | Current frame # | Mark the first index of data array |
| 3 | Total frame # | Mark the lengh of data array |
| 4 | None | Reserve |
| 5 | None | Reserve |
| 6 | Column parity | Column parity of current data frame, enhance robustness |
| 7 | Row parity | Combine current data frame and column parity, calculate parity, double enhance robustness |

Once we finished the construction of the data array and label array for each data frame, just combine these two arrays horizontally into one array, which we named as Information Array, we can obtain the array filled of information. Figure 8 shows the structure of information array.

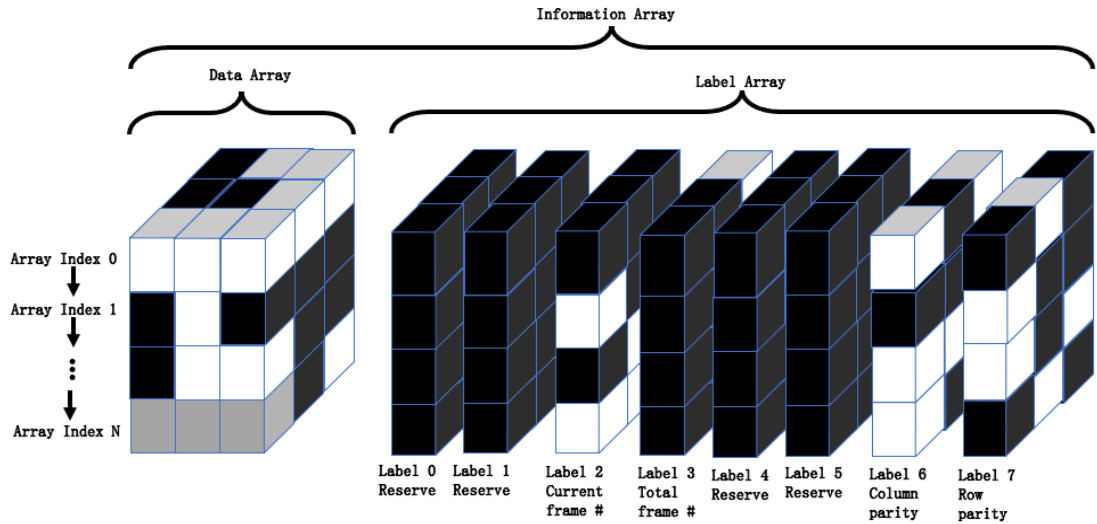

Fig. 8: Combine data array and label array horizontally to information array, each data frame corresponding a label frame, identified by array first index. One small cube represents one value, a black cube represents 0, a white cube represents 255, a gray cube represents 128.

## d) Scaling Up Each Pixel to Block

Pixel is a very small and basic vision unit on screen, which is very often got lost or damaged on network transmission and video compression. Normally, human vision can not identify it and nothing impacted. But to operation system, pixel damaged means data damaged, that will cause data recovry fail.

So it is necessary to add some redundant data in case of data damaged. By scaling one cube to 5x5 cube block, for example, one white pixel is a cube with value 255, will be scaled to 5x5 cube block with all values 255, and a frame with size NxM cubes will be scaled to 5Nx5M array.

By this operation, we add redundant data into array to against pixels damaged. While decoding process, it is easy to locate each cube block and calculate mean value to recover corrrect value even some pixels are damaged.

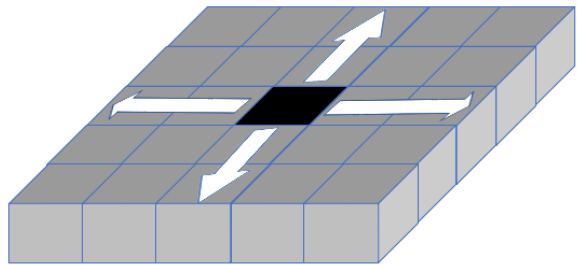

Fig. 9: Scale one cube to 5x5 cube block with same value to enhance redundancy in case of some cubes damaged.

## e) Constructing Container Array

When recording screen, user needs to draw a rectangle to asign recording region, which should be media player window. It has to expand region border to contain all necessary information and avoid data loss. So the additional problem is to identify the useful pixels from each video frame.

The solution is packing information array into a container array. As figure 10, container array is a 3D array which has black external wall with value 0 and white internal wall with value 255, information array is installed into it. By the structure, in decoding process, we can locate each container frame from video frame clearly and precisely with simple computer vision skill plus some rules, here we use findContours API in OpenCV3.x/4.x to find some rectangles, and filter out the container rectangle by area, turns out it is works fine.

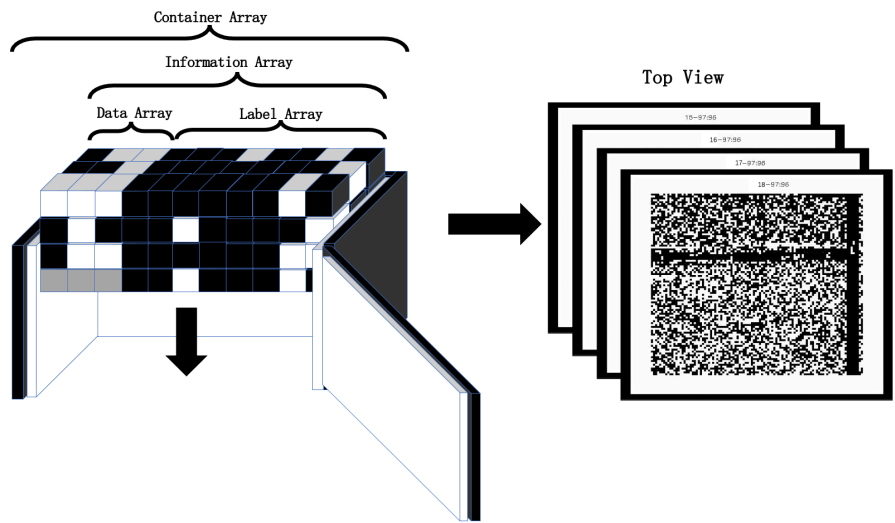

Fig. 10: Construct container array, put information array into container array to generate final video.

## f) Generating Video File

After constructed the whole array with all necessary information, the final step is to generate video file by OpenCV. OpenCV is a widely used computer vision library supporting many types of video file, here we choose MP4 which is widely supporting by most of video players.

Considering the container array is a 3D array, we can write each video frame with each 3D array frame in sequence by first array index.

Some player like MS media player will drop some frames in the first 1~2 seconds while playing video, maybe relate to cache algorithm, that does not impact human watching video but does seriously impact data integrity.

To resolve this issue, we just repeat servral times while writing the first frames to video file. Further more, considering random frames would be lost in network transmission, to repeat the whole array two or more times while writing video file for better robustness, here we repeat two times default.

## 3.2 Decoding Process

Technically, once we played the video in VDI and recorded it in host PC(described in Section 2: Operation Model), the full information of the target file has been pulled out of VDI, only with human eyes can not recognize them, but CyberEye can.

Basically, decoding process is a reversed encoding process, which is pretty simple because of the preparations of the encoding process. Still, there are some tricks in the process.

## a) Denoising and Isolating the Container Array

As the previous section 2.3, user can choose any screen recorder tool and draw a rectangle window to define recording region while video playing. The operation is simple but it's involving many unnecessary pixels and these pixels value may be changed by most of media lossy compression algrithms. For each frame, the noise must be reduced and standarized to avoid value faults and array border line shifting.

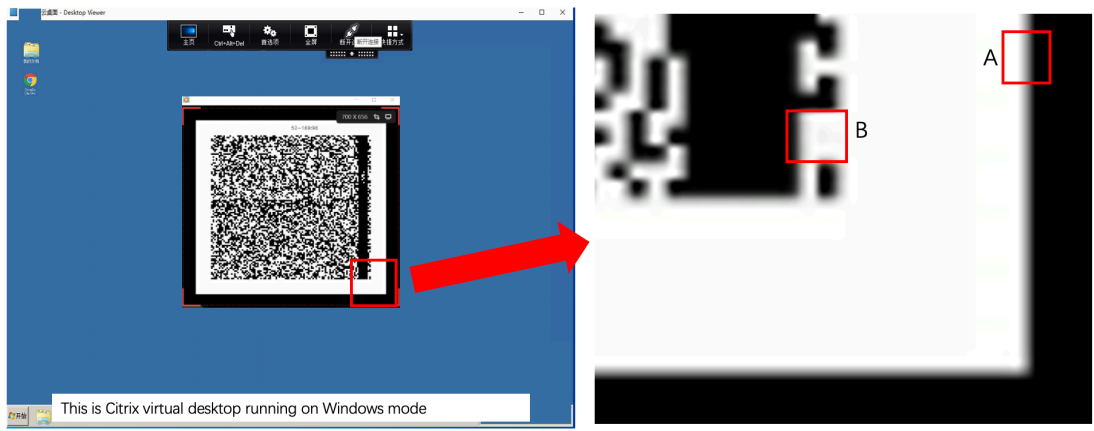

Fig. 11: A and B shows image artifact by media compression algrithms cause frame blur. Two or more pixels transition near the border, for example, value 0 destabilized to 17, value 255 destabilized to 248.

To smooth discrete noise pixels, OpenCV medianBlur API is quick and effective, after that, perform binaryzation operation by correcting pixel value to 0 if value less than or equal 127, to 255 if value greater than 127, here 0 means black pixel, 255 means white pixel. After we got smooth and pure binary frame, with OpenCV findContours API, we can isolate the container region easily and precisely.

## b) Mean-pooling and Correcting Bits Value

By the last step, we have isolated the region of container array precisely frame by frame, with the preparation described in Section 3.1e, we can easily strip the wall and extract data and label array since we know the thickness of the wall and the size of the arrays.

But there still exist some offset pixel values caused by media compression algorithm while video generation and recording, such as origin pixel value 255 offset to 247, and 0 offset to 18, 128 to 121, there are random offsets in range of 3~30 in survey.

It is why we scale each pixel to a block as introduced in Section 3.1d, it will increase additional robustness into array. Even some random pixel values offset, but with calculating mean-pooling of each block, we can still estimate the most likely value, all these values are around ground true value 0, 255 and 128 with small distance, we just correct these estimated values to the nearest ground true values, we can get a more precise array. Then reset all 255 to 1, all 128 to 2 and maintain 0 to 0, we can get the standardized information array filled with value 0, 1, 2. Note value 2 should be in array bottom exclusively as data end signal.

## c) Parity Check and Deduplicate Array

There are still very few wrong values and many duplicate data frames, because FPS of original video is always less than the recorded video. In the experiment, set 5 FPS to original video will balance video durition and data integrity, set greater than 25 FPS to recoding video will keep all useful information. In MacOS, Quicktime default works on 60 FPS, and in Windows, most screen recording tools default work on 25~30 FPS, all works fine, but higher FPS is better in case of the frames lost.

Though there will be many duplicate data frames, with the preparation described in Section 3.1c and 3.1d, just perform partity check to each data frame, drop false data frames and deduplicate remaining data frames. Then, check up to make sure that all necessary data frames are existed and passed, if true, to rebuild the final data array, which should contain all and precise original information, no more, no less.

### d) Reshaping Array and Recovering File

In Section 3.1a, we reshape a bits sequence to array, now we just reverse this process, reshape data array to 1D bits sequence, cut off the sequence after the first stop signal, then convert binary bits to hexadecimal, and save to file. Finally, we recover the original file from video precisely and successfully.

# 4. Evaluation

This section provides an evaluation of the practice with the approach we introduced above, and because Citrix VDI is the most popular virtual desktop solution used in enterprises, we choose it to show the effectiveness and performance.

**Environment**:

Host PC: MacBook Pro (Retina, 15-inch, Mid 2015), 2.2 GHz Intel Core 7, 16GB DDR3, Intel Iris Pro 1536, MacOS 11.6 with internet access, bandwidth 50Mb, locate at city A.

Recording application in host PC: QuickTime Player 10.5(1086.4.2), preinstalled in MacOS.

VDI Client installed in host PC: Citrix Workspace Version 21.12.0.32(2112), virtual desktop is Windows Server 2008 R2 Enterprise(version 6.1.7601, sp1), with preinstalled Windows Media Player.

VDI data center is located at city B, around 2,000 kilometers distance away from city A.

CyberEye is developed with Python 3.7.3 with main dependencies on OpenCV 3.4.2 and Numpy 1.21.5, compiled excutable version with Pyinstaller 5.3 supporting Win 2008 and Win 7.

We evaluated different size of target files to verify the effictiveness and performance, note we only use ZIP type file as the target file because all types of files can be archieved to this type. Encoding file type is MP4, recording file type is MOV from QuickTime on 60 FPS, array repeated twice. Note QuickTime will interrupt the last frame while writing video file, which cause fail when decoding the file, use FFMPEG application to fix it by the following command, and any other media tools would work also. About FFMPEG, refer: https://github.com/FFmpeg/FFmpeg.
The easy fix command is: ffmpeg -i recorded_video.mov -vcodec copy recorded_video_fix.mp4.

Table 2: Encoding and Decoding Performance

| ZIP File Size (KBytes) | Encoding Time (Seconds) | Encoded Video Size (KBytes) | Encoded Video Length (Seconds) | Recorded MOV Video Size (KBytes) | Decode Time (Seconds) |
| --- | --- | --- | --- | --- | --- |
| 128 | 2.231 | 37,182 | 67 | 140,752 | 76.338 |
| 256 | 3.627 | 64,389 | 105 | 229,110 | 159.342 |
| 513 | 6.049 | 118,789 | 194 | 383,230 | 293.832 |
| 1,025 | 11.04 | 227,866 | 372 | 829,678 | 561.398 |
| 2,051 | 21.149 | 445,825 | 728 | 1,544,364 | 1,098.988 |
| 3,077 | 32.757 | 663,943 | 1085 | 2,172,970 | 1,634.899 |
| 5,128 | 51.072 | 1,099,703 | 1,797 | 3,855,984 | 2,699.382 |

Through above practice, we can validate the approximate linear relationship between ZIP file and video size, including the encoding and decoding time. The original ZIP files are encoded by 5 FPS, repeating twice, and recorded video are default running on 60 FPS by QuickTime. From table 2, 1,025KB ZIP file will be encoded to 829MB video file with 372 seconds length, greater file will cost greater disk space and time. For reliability reason, we suggest to split large ZIP fille to smaller files less than 1MB.

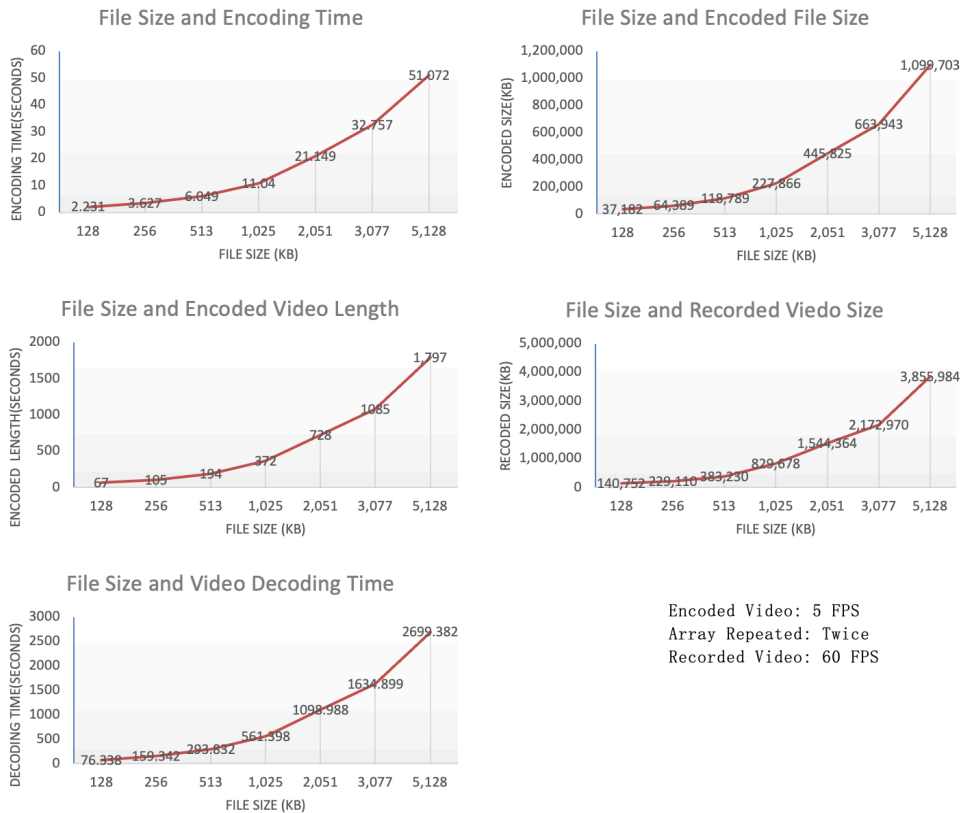

Fig. 12: Performance evaluation by ZIP file size, approximate linear relationship between ZIP file and video size, including the encoding and decoding time.

# 5. Conclusion

This paper shows a new data exfiltrate approach called CyberEye, to transmit data out of restrictly virtual desktop via playing and recording video, even data transmission out of virtual desktop has been forbidden by VDI policy.

We developed and opened source code of CyberEye, which can encode any file in virtual desktop to a video, to play and record it in host PC, then decode video for restoring original data file. CyberEye is a green software developed by Python3 with dependencies OpenCV and Numpy, compiled to excutable version running on Windows OS, which only requires the same permissions with current authorized user because all operations are smooth and safe.

We've presented the operation model and provided the related technical works, also have opened the source code and demostration to enable researchers to reproduce the results. We've introduced the details of CyberEye about the design and implementation, and have evaluated CyberEye on Citrix VDI to show data can be exfiltrated successfully from virtual desktop with approximate linear relationship between target file size and other efforts.

CyberEye is a new data exfiltration approach using computer vision, which means even it is visible to human but is still invisible to the traditional antivirus and firewall applications, so it is undetectable and does not trigger alarms.

According to this approach based on computer vision, we should prove CyberEar based on ultrasonic wave in future work, which is invisible and inaudible to both human and safe guard applications.

https://ieeexplore.ieee.org/document/6975588.

**Some toolkits we used in this paper:**

1. OpenCV: A pupular and powerful computer vision toolkits. Refer: https://opencv.org/

2. Numpy: The fundamental package for scientific computing with Python. Refer: https://numpy.org/

3. FFMPEG: A complete, cross-platform solution to record, convert and stream audio and video. Refer: http://www.ffmpeg.org/

4. Apower: Cross-platform Screen Recorder. Refer: https://www.apowersoft.com/

5. Citrix: https://www.citrix.com/